# Low-power, agile electro-optic frequency comb spectrometer for integrated sensors

Kyunghun Han,[1,2,3,†] David A. Long,[1,†,*] Sean M. Bresler,[1,2] Junyeob Song,[1,3] Yiliang Bao,[1,3] Benjamin J. Reschovsky,[1] Kartik Srinivasan,[1,2] Jason J. Gorman,[1] Vladimir A. Aksyuk,[1,*] and Thomas W. LeBrun[1]

[1]*National Institute of Standards and Technology, Gaithersburg, Maryland 20899, USA*
[2]*University of Maryland, College Park, Maryland 20742, USA*
[3]*Theiss Research, La Jolla, California 92037, USA*

[†]*These authors contributed equally.*

*\*david.long@nist.gov* (D. A. Long) and *vladimir.aksyuk@nist.gov* (V. A. Aksyuk)

Sensing platforms based upon photonic integrated circuits have shown considerable promise, however they require corresponding advancements in integrated optical readout technologies. Here, we present an on-chip spectrometer that leverages an integrated thin-film lithium niobate modulator to produce a frequency-agile electro-optic frequency comb for interrogating chip-scale temperature and acceleration sensors. The chirped comb process allows for ultralow radiofrequency drive voltages, which are as much as seven orders of magnitude less than the lowest found in the literature and are generated using a chip-scale, microcontroller-driven direct digital synthesizer. The on-chip comb spectrometer is able to simultaneously interrogate both the on-chip temperature sensor and an off-chip, microfabricated optomechanical accelerometer with cutting-edge sensitivities of $\approx 5$ µK·Hz$^{-1/2}$ and $\approx 130$ µm·s$^{-2}$·Hz$^{-1/2}$, respectively. This platform is compatible with a broad range of existing photonic integrated circuit technologies, where its combination of frequency agility and ultralow radiofrequency power requirements are expected to have applications in fields such as quantum science and optical computing.

Photonic integrated circuit (PIC) technologies hold tremendous potential for low cost, high accuracy field-deployable sensing. However, unlocking these capabilities requires chip-scale integration of not only the sensors but also the optical readout. Chip-scale optical frequency combs are well suited to these photonic readout demands due to their capability for high speed, multiplexed measurements without the need for any moving parts, [1] thus allowing for transduction of the photonic sensor to a digital output. In particular, electro-optic frequency combs can not only be integrated, but also are capable of having sufficient frequency agility to achieve the high resolution required to probe atomic transitions as well as optical (and optomechanical) cavity-based sensors, where a measurement of the cavity motion is required to read out the sensor. [2,3] These type of measurements typically require narrow comb tooth spacings at the MHz level and comb spans at the GHz level, leading to a sensitive and high dynamic range readout.

Recent advancements in thin-film lithium niobate (TFLN) technology have enabled the development of compact on-chip electro-optic modulators (EOMs) with > 100 GHz modulation bandwidth [4–7] and half-wave voltages, $V_\pi$, which outperform traditional bulk lithium niobate EOMs. [8] When electro-optic frequency comb spectroscopy is integrated with on-chip EOMs, it becomes a scalable platform for multiplexed solid-state on-chip spectroscopy [9–17]. However, these previous integrated electro-optic comb generation approaches have relied upon high radiofrequency drive powers between 0.3 W to 4 W (25 dBm to 36 dBm) and wide comb tooth spacings (generally in the GHz) which restrict their application for sensing applications.

Here we present an on-chip integrated spectrometer that allows for frequency-agile, electro-optic frequency combs generated by ultralow power radiofrequency waveforms (25 nW to 4 mW, i.e., −46 dBm to 6 dBm) which are readily compatible with even the lowest available on-chip powers. These combs are generated through the use of a chirped waveform [18,19], rather than the traditional single tone sinusoid, leading to hundreds or thousands of flat comb teeth even at these ultralow powers. We note that this approach is not specific to the present device or modulator but rather is applicable to any non-resonant electro-optic modulator.

The use of serrodyne modulation to shift the local oscillator [20] (in place of the traditional acousto-optic modulator) greatly reduces the fabrication complexity. Further, the fabrication approach presented herein allows for compatibility with a wide range of sensors based on silicon nitride PICs on the same chip as well as off-chip sensors. As a result, we have concurrently probed a temperature sensor integrated on the same chip as the electro-optic modulators as well as a separate fiber-connectorized chip-scale optomechanical accelerometer. With both sensors we achieve state-of-the-art [21–24] performance with temperature and acceleration sensitivities of $\approx 5$ µK·Hz$^{-1/2}$ and $\approx 130$ µm·s$^{-2}$·Hz$^{-1/2}$ ($\approx 13$ µg/Hz$^{1/2}$, where 1 g = 9.80665 m/s$^2$), respectively, demonstrating the power of the integrated, frequency-agile optical frequency comb spectrometer.

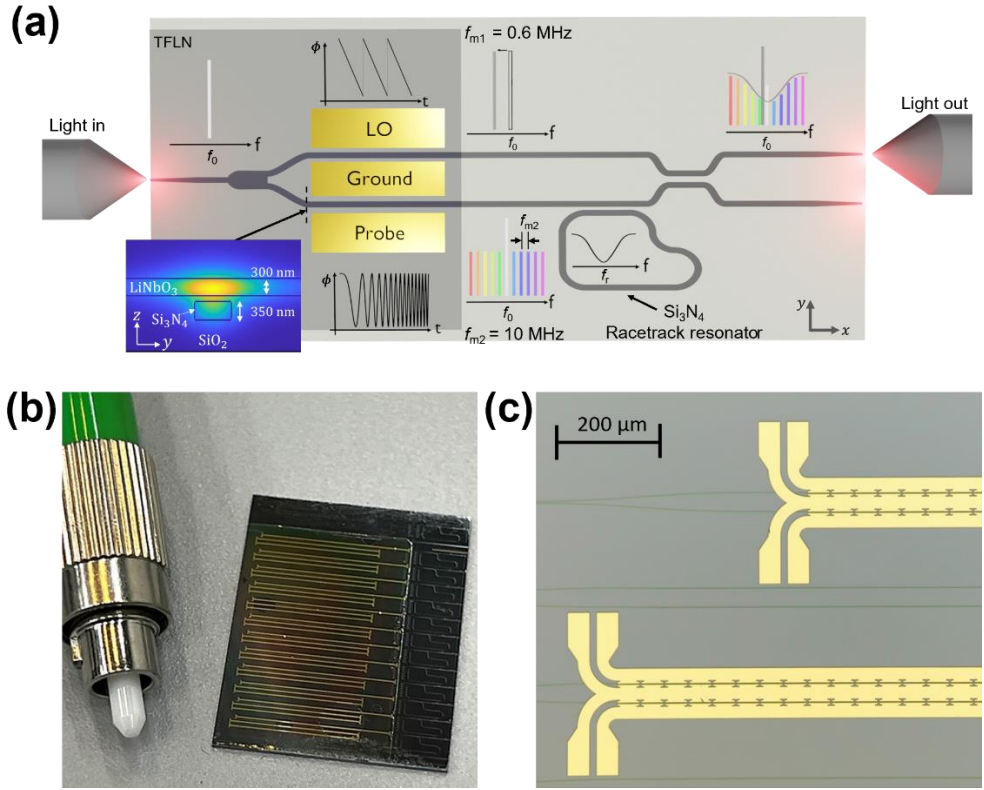

Fig. 1. **a.** Schematic of the integrated electro-optic frequency comb spectrometer. The left portion of the chip (shaded area) is a bonded thin-film lithium niobate (TFLN) layer for active modulation. Input light, injected from the left lensed fiber with a carrier frequency, $f_0$, is split into two arms. The lower arm (labeled as Probe) is modulated by a chirped waveform, generating an electro-optic frequency comb with a repetition rate of $f_{m2} = 10$ MHz. This comb then interrogates a given resonance, located at $f_r$, of the silicon nitride racetrack resonator. The upper arm, which serves a local oscillator (LO), is modulated by a sawtooth waveform, shifting its center frequency by $f_{m1} = 600$ kHz through the serrodyne process. The probe and LO are then combined before being sent to an off-chip photodetector. The inset shows a cross-sectional view of the waveguide with the electric field intensity distribution. **b.** Photograph of an integrated

electro-optic frequency comb spectrometer chip containing twenty-four individual electro-optic modulators and integrated temperature sensors. A fiber ferrule is shown for scale. **c.** Optical microscopy image of the electrode area. The capacitively loaded slow wave electrode design was used for $V_\pi$ and bandwidth measurement. The conventional unpatterned electrode design was used for the photonic readout of the two cavity-based sensors.

The TFLN-based electro-optic frequency comb spectrometer is composed of a Mach-Zehnder interferometer (MZI) with an electro-optic phase modulator on each leg (Fig. 1). The light from the lensed fiber is coupled to the input waveguide, consisting of a silicon nitride ridge waveguide with a width of ≈ 2.5 µm and a lithium niobate slab layer vertically spaced by ≈100 nm. The input light is divided into two arms by a multi-mode interference 1×2 splitter. The waveguide width becomes narrower (≈ 650 nm) after the splitter to efficiently modulate the phases by expanding the mode into the lithium niobate layer. The electro-optic modulator in the lower arm is driven by a chirped waveform [18] to generate the electro-optic frequency comb, while the upper modulator receives a serrodyne modulation to generate a frequency-shifted local oscillator for heterodyne detection [20]. The frequency comb signal is first used to interrogate the on-chip silicon nitride racetrack resonator, which acts as a temperature sensor. Subsequently the optical signals are combined via a symmetric 2×2 directional coupler and sent via an output optical fiber to probe the Fabry-Pérot optical cavity of the physically separate chip-scale optomechanical accelerometer before being detected. A more detailed schematic and description of the optical layout can be found in the Supplementary Material.

The optical waveguides are fabricated from stoichiometric silicon nitride with a thickness of ≈ 350 nm. All features are patterned by 365 nm ultraviolet stepper lithography to allow for low-cost mass-production. A ≈ 300 nm thick, X-cut TFLN layer is bonded on top of the waveguide with an ≈ 100 nm silicon dioxide gap. Approximately 4 cm$^2$ TFLN squares are bonded to the PIC wafer via direct bonding of lithium niobate on insulator chips with subsequent Si handle layer removal by mechanical polishing and reactive ion etching (see Methods for further details). The TFLN layer is bonded selectively where active phase modulation is required.

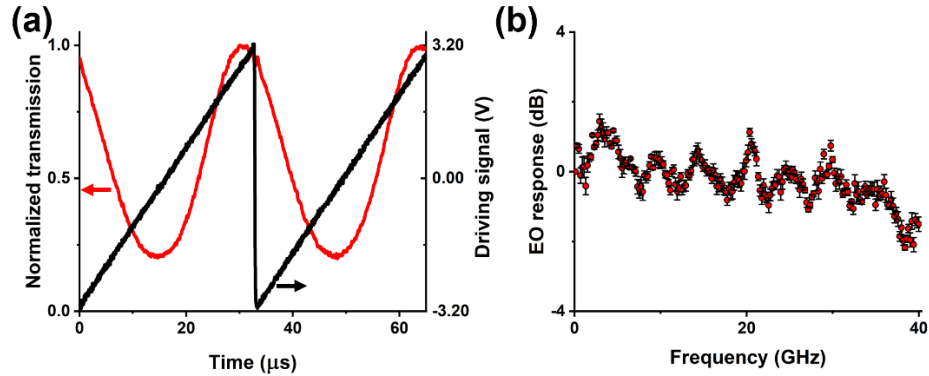

Fig. 2. Performance of the fabricated devices. **a.** Half-wave voltage, $V_\pi$, measurement. The normalized transmission through the interferometer (blue, left axis) approaches a near-continuous sine wave when driven with a 30 kHz triangular signal (red, right axis) at $V_\pi$ ≈ 3.2 V. **b.** Average electro-optic response of the fabricated EOM from five repeated measurements, yielding a 3 dB bandwidth of greater than 40 GHz. The shown uncertainties are the standard error of the mean from the repeated measurements.

The individual TFLN EOMs (see Fig. 1c) were characterized by applying a 30 kHz triangular waveform to only one leg, while the phase of the other leg remained unmodulated (see Fig. 2a). The measured $V_\pi$ was ≈ 3.2 V in a modulation length of ≈ 1.2 cm, equivalent to a half-wave voltage - length product, $V_\pi L$, of ≈ 3.84 V·cm. This value is comparable to recent work on heterogeneously integrated EOMs. [25–27] The $V_\pi L$ is fundamentally influenced by the dimensions of the electrode gap, with a preference for minimizing this gap to enhance modulator performance. However, this reduction in gap size inherently leads to increased optical losses due to the proximity of the metallic electrode. To mitigate such losses, the introduction of a thin silicon dioxide ($SiO_2$) layer between the lithium niobate layer and the metal electrode has been identified as an effective strategy [28]. This $SiO_2$ layer acts as a buffer, reducing the interaction between the optical field and the electrode, enabling closer electrode spacing. We note that the drive voltage could be further reduced by a factor close to two by driving the modulator in a push-pull configuration.

In the context of direct electro-optic comb spectroscopy, the operational bandwidth is constrained by the maximum frequency achievable in the chirp signal, typically around a few gigahertz. Consequently, we tailored the electrode design to achieve a reduced $V_\pi L$, though this results in a trade-off with a diminished modulation bandwidth. The electro-optic response of the modulator was measured by sweeping the modulation frequency and measuring the RF power from the photoreceiver at the modulation frequency with the RF spectrum analyzer. This led to a 3 dB bandwidth of >40 GHz (see Fig. 2b). The measurement set-up was calibrated by placing the input and output RF probe side-by-side without the device-under-test on top of the insulating surface. The capacitively loaded slow wave electrode design is the same as in previous research with a silicon substrate [29] except for the electrode gap size. The low extinction ratio is designed to allow for higher optical power in the electro-optic comb leg. The oscillation in the frequency response originates from the characteristic impedance mismatch of the coplanar waveguide. The free-spectral range of the oscillation is close to the estimated value given by the product of the electrode length and the RF phase velocity. This oscillation has a minimal impact on the final spectroscopy signal since any residual distortion is normalized out by measuring a reference spectrum in the absence of the device-under-test as explained in the supplementary material.

For EOM comb generation, the lower modulator was driven by a time-periodic frequency chirp voltage $V = A \sin\left[2\pi(f_1 + \frac{(f_2-f_1)t}{2\Delta t})t + \varphi\right], 0 < t < \Delta t$, where $A$ is the amplitude, $f_1$ and $f_2$ are the initial and final chirp frequencies, respectively, $\Delta t$ is the time duration of the chirp, and $\varphi$ is a constant phase [18]. This results in a frequency comb whose bandwidth is $2(f_2 - f_1)$ and a repetition rate of $1/\Delta t$. The frequency chirp was produced via a chip-scale, direct digital synthesizer which was controlled by a microcontroller, allowing for frequency agile, digital control over the comb parameters [30]. For the present measurements we utilized a comb having a bandwidth of 3 GHz with a 10 MHz repetition rate (i.e., 300 teeth in the comb).

Critically, because the EOM is driven by a radiofrequency comb, rather than relying upon nonlinearities in the EOM itself, the number of observed comb teeth is independent of the applied radiofrequency power. As a result, we can readily observe optical frequency combs with radiofrequency drive powers as low as 25 nW (−46 dBm) (see Fig. 3a). This drive power is ≈70 dB lower than the lowest drive voltage we are aware of for an integrated electro-optic frequency comb (see Fig. 3b). These ultralow powers are particularly well suited to the low logic levels of complementary metal-oxide-semiconductor (CMOS) and Gunning transceiver logic (GTL), thus precluding the need for high power, off-chip amplifiers. For the remainder of this paper we used a radiofrequency drive power of 4 mW (6 dBm).

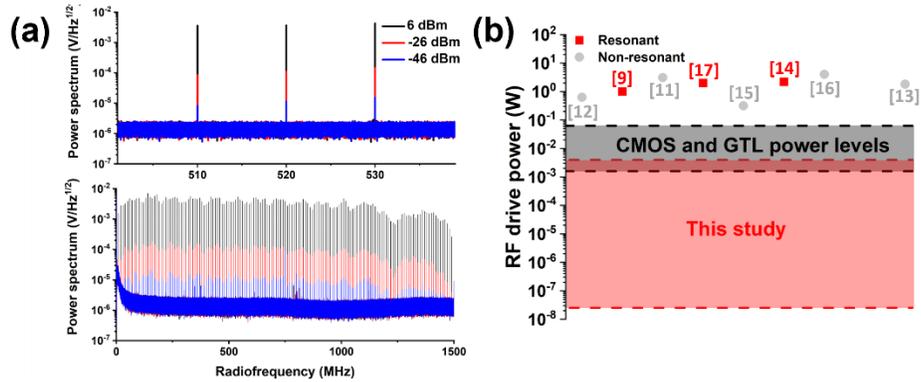

Fig. 3. **a.** Photodetected power spectrum of the integrated optical frequency comb for a range of root-mean-square radiofrequency (RF) chirp powers. Comb teeth are clearly visible even at drive powers as low as 25 nW (−46 dBm). Each of the shown power spectra is the average of one hundred spectra, each of which was acquired in 10 ms. **b.** Comparison of the RF drive powers used in this study to those in the literature for both non-resonant (i.e., frequency-agile) and resonant (essentially fixed frequency) devices. The literature values are indicated by their corresponding reference numbers. We note the lowest RF drive power used in the present study is 71 dB lower than the lowest previously reported value while still maintaining frequency agility. Also shown are the corresponding ranges for CMOS and GTL logic levels.

After the EOM comb is generated, it is passed through the TFLN bonding boundary and into a silicon nitride racetrack microresonator with a nominal length of 7.80 mm and a measured loaded quality factor of ≈$2.6\times10^5$ (see Fig. 1a). A mode expansion simulation using the known thermo-optic coefficients of silicon nitride and silicon dioxide [31] predicts a thermal shift for a given optical cavity mode of ≈2.5 GHz/°C, allowing the racetrack resonator to serve as a sensitive, integrated temperature sensor.

In order to down-convert the optical frequency comb into the radiofrequency domain for digitization via self-heterodyne detection, a local oscillator (LO) that is frequency shifted with respect to the initial laser frequency is required. This frequency shift translates the center of the radiofrequency comb away from DC and ensures that the comb teeth occur at unique radiofrequencies. Traditionally in fiber-optic comb systems this LO frequency shift is achieved with an acousto-optic modulator (AOM). However, while integrated acousto-optic modulators based on TFLN have been demonstrated [32], the complexity of incorporating acousto-optic devices onto the same platform as our on-chip EOMs and temperature sensors - without degrading the performance of any element - suggests that a pure electro-optic approach would be advantageous. As a result, here we have utilized serrodyne electro-optic phase modulation [20] to produce the required frequency shift. Serrodyne modulation was produced by a 600 kHz sawtooth waveform from a commercial function generator. We note that this benchtop function generator could be replaced by more compact solutions such as commercially available, chip-based function generators.

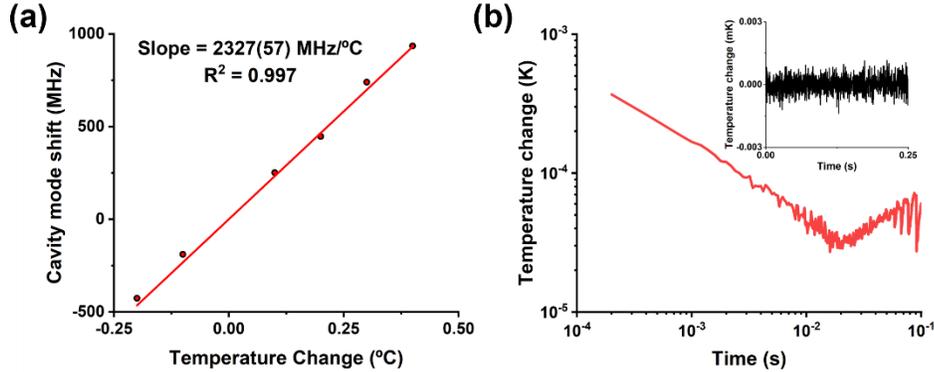

Fig. 4. On-chip racetrack resonator temperature dependence. **a.** Measured cavity mode shift as a function of temperature as well as a linear fit. Each data point is the average of 500 individual Lorentzian fits to 200 µs long measurements. Temperature and frequency change is referenced to measurements made at ≈ 27.1 °C. The standard error of the mean for each frequency data set was between 36 kHz and 93 kHz, leading to statistical uncertainties much smaller than the shown data point size. We observe good linearity in these measurements with the shown slope being near the calculated value of ≈ 2500 MHz/°C. **b.** Allan deviation of repeated measurements of the chip's temperature made by measuring the cavity's mode frequency and converting to temperature based on the slope determined in panel **a**. At ≈ 20 ms we reach a minimum of the Allan deviation of 30 µK. The inset gives the corresponding temperature change time series.

After the optical frequency comb has probed the evanescently coupled racetrack microresonator, it is combined with the serrodyne-shifted LO by a 2×2 directional coupler. A lensed fiber was employed to couple the light off chip. For the measurements in the absence of the optomechanical accelerometer, the optical power was detected by a photodiode. The resulting interferogram was digitized, Fourier transformed, and normalized, producing an optical frequency comb spectrum. An integrated resistive heater driven by a proportional-integral-derivative servo was employed to control the chip's temperature. As can be seen in Figure 4a, as the temperature of the chip was changed, the integrated optical frequency comb spectrometer was readily able to measure the shift of the microresonator optical cavity mode, allowing for a direct measurement of the resonator's temperature. An Allan deviation measurement, performed while the chip's temperature was actively controlled with the resistive heater to slightly above room temperature, reached a minimum of ≈30 µK (see Fig. 4b). We note that this noise floor lies below that of cutting-edge photonic thermometers (e.g., Ref. [21] which achieved 70 µK). From this Allan deviation we can extract the temperature sensitivity as ≈5 µK/Hz$^{1/2}$ for measurement durations <20 ms. No optical power induced shift of the resonance was observed, even when ≈20 mW was injected on chip (see Supplementary Material).

Unlike laser locking approaches, the integrated comb spectrometer presented herein provides full spectrum information for each measurement and can be used to dynamically extract features of the cavity mode shape, such as coupling efficiency and finesse for multiple devices simultaneously. As a demonstration, we have performed simultaneous measurements of the on-chip temperature sensor and a chip-scale optomechanical accelerometer located off-chip. The fiber-connectorized optomechanical accelerometer is based upon a plano-concave Fabry-Pérot cavity that is ≈200 µm long and is composed of a static, microfabricated concave mirror (radius of curvature of ≈270 µm) and a planar mirror fabricated on a movable proof mass suspended by silicon nitride beams (see Fig. 5a and 5b) [33]. The translation of the proof mass due to acceleration produces a shift of the cavity resonance. The accelerometer employed in these measurements had an optical finesse of ≈9200, a mechanical quality factor of 115(1) and a mechanical resonance frequency of 23.686(1) kHz, where the shown uncertainties are one standard deviation in the fit parameters for the thermomechanical noise spectrum. For these measurements, the accelerometer was mounted on a closed-loop electromechanical shaker table with a commercial reference accelerometer to apply a known acceleration.

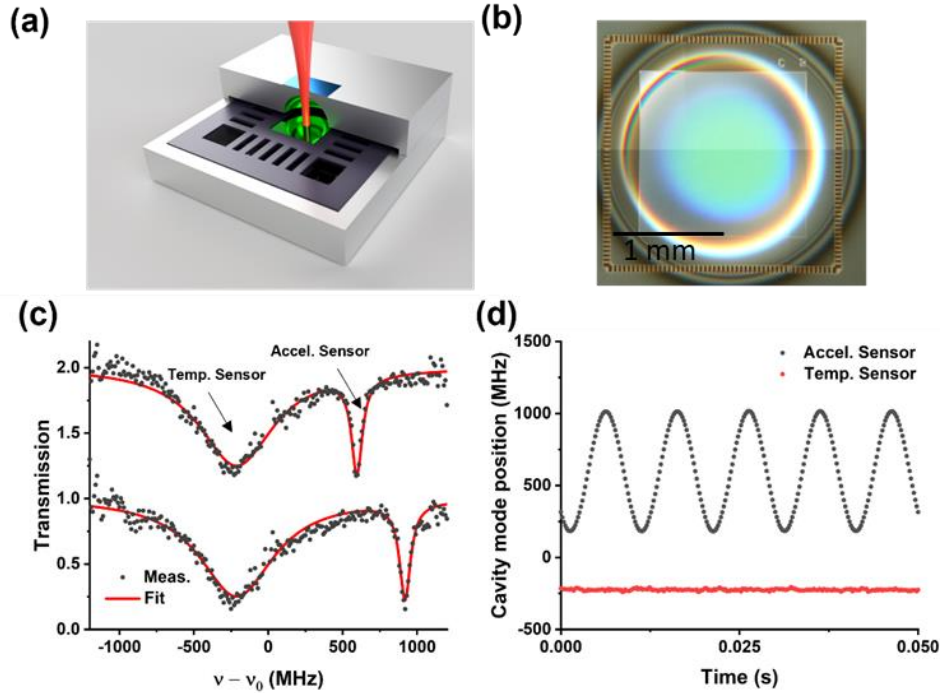

Fig. 5. **a**. Schematic of the chip-scale optomechanical accelerometer (shown not to scale). The two bonded silicon chips are shown in white with the silicon nitride beams shown in dark gray. The highly reflective mirror coatings are shown in green with the anti-reflective coatings shown in blue. In the presence of acceleration, the plano-concave cavity's length will change due to the motion of the proof mass to which the planar mirror is attached. This cavity mode displacement can be readily recorded with the on-chip optical frequency comb spectrometer (with the light shown in red). **b**. A stitched optical microscope image of the accelerometer showing the proof mass resonator with a planar mirror (at center) and supporting silicon nitride beams around the edge of proof mass. **c**. Simultaneously measured spectra of the on-chip racetrack-resonator-based temperature sensor (Temp. Sensor) and the chip-scale optomechanical acceleration sensor (Accel. Sensor). Two separate comb spectra are shown for two separate moments in time, with the upper spectrum vertically offset by 1. While the temperature sensor resonance is nominally static, the accelerometer resonance translates in the presence of acceleration. The shown comb spectra were recorded in only 200 µs with no averaging performed. The x-axis is relative to the measured wavemeter frequency of $v_0 = 191\ 497\ 139$ MHz. Also shown are two corresponding Lorentzian fits. **d**. Retrieved cavity mode positions for the two sensors when the Accel. Sensor was driven by a 9.81 m/s$^2$ (i.e., 1 g) acceleration at a frequency of 100 Hz. Each of the shown data points was recorded in 200 µs.

For the simultaneous measurements of the two chip-scale sensors, after the optical frequency comb had probed the racetrack resonator and been combined with the LO, the light was sent via fiber to the optomechanical accelerometer through a circulator, with the circulator return connected to the photodiode. The resulting comb spectrum of the two chip-scale cavity-based sensors can be found in Fig. 5c, with the broader resonance arising from the racetrack-microresonator-based temperature sensor and the narrower resonance due to the optomechanical acceleration sensor. Based upon the standard deviation of repeated measurements made in the absence of an external acceleration, the displacement sensitivity of the accelerometer when interrogated by the integrated comb spectrometer was determined to be ≈4 fm/Hz$^{1/2}$ at a measurement rate of 200 kHz. Because of the large separation between mechanical modes of the optomechanical accelerometer (no other modes below 100 kHz), we can readily invert the measured cavity mode displacements to yield acceleration [34]. This led to an acceleration sensitivity of ≈130 µm·s$^{-2}$·Hz$^{-1/2}$ (≈13 µg/Hz$^{1/2}$ where 1 g = 9.80665 m/s$^2$). Despite the fact that the present measurement utilized an integrated optical readout and was also simultaneously measuring the temperature sensor, our acceleration sensitivity is similar to that achieved in other optomechanical accelerometers (7.8 µg/Hz$^{1/2}$ to 10 µg/Hz$^{1/2}$) [22–24]

When the electromechanical shaker table was excited at 100 Hz, we observed clear motion of the optomechanical accelerometer's cavity mode, while, as expected, the racetrack resonator exhibits only slow motion due to thermal drifts (see Fig. 5d). Thus, the integrated comb spectrometer can readily interrogate these two sensors simultaneously with minimal crosstalk. We note that the data presented in Fig. 5d was recorded with a time resolution of 200 µs and that time resolution as short as 5 µs is possible with the presented comb settings. For the measurement shown in Fig. 5d, this yielded a peak acceleration of 9.75(2) m/s$^2$ which is within 0.56 % of the value reported by the reference accelerometer. We note that this is excellent agreement given the combined one standard deviation uncertainty of the reference accelerometer is 0.5 %. Further, this matches the highest level of accuracy which has been demonstrated for an optomechanical accelerometer. [35] This is especially notable given that the present approach utilized an integrated comb spectrometer rather than bulk optics and that the present system was also simultaneously reading out the temperature sensor.

Here we have demonstrated an integrated frequency-agile, electro-optic frequency comb platform which is readily compatible with silicon nitride photonic integrated circuits as well as CMOS and GTL powers. The frequency agility of this platform make it ideally suited for photonic readout of optomechanical sensors as well as the narrow atomic features found in quantum systems. In addition, the capability to interrogate multiple unique sensors with a single integrated photonic readout represents an important step toward fully integrated systems such as inertial navigation units and quantum sensors. Further, we note that the present measurements made use of only a single MZI, with each of the present chips containing twelve MZIs with twelve separate temperature sensors. As a result, the present approach is highly scalable, with >900 MZIs being capable of being fabricated on a single 100 mm (4″) wafer. This high degree of scalability as well as the flexibility of this method make it applicable in wide ranging fields such as quantum sensing, optical computing, and cavity optomechanics.

**Methods**

*Fabrication*

A stoichiometric silicon nitride layer with a nominal thickness of 350 nm was deposited by low-pressure chemical vapor deposition (LPCVD) on top of a thermally oxidized silicon dioxide layer with a nominal thickness of 3 µm as a buried oxide layer (BOX). The waveguide was patterned by a UV lithography stepper with a wavelength of 365 nm and reactive ion-etching with a CF$_4$/O$_2$ gas mixture. After removing the photoresist with a piranha cleaning solution, a plasma-enhanced chemical vapor deposition (PECVD) tool was used to deposit nominally 1 µm of silicon dioxide. The corrugated top surface of the wafer was planarized by a chemical-mechanical polishing (CMP) tool, leaving ≈100 nm of silicon dioxide on top of the silicon nitride waveguide. Aluminum oxide layers (≈3 nm thickness) were coated on both the silicon nitride waveguide wafer and the diced lithium niobate on insulator (LNOI) wafer. After a deionized wafer spray cleaning, the two samples were held together for wafer bonding at room temperature and atmospheric pressure. The bonded wafer was annealed at 200 °C for 1 hour to enhance the bonding strength. Next, a ≈3 µm thick silicon dioxide layer was deposited by the PECVD tool to act as a protecting layer for the silicon nitride waveguide circuit in regions without the TFLN. The silicon substrate of the bonded LNOI piece was then removed by mechanical polishing, leaving ≈50 µm of silicon, with the rest of the substrate removed by reactive ion-etching with $SF_6$. The BOX layer of the LNOI piece was then thinned with a buffered oxide etcher. The final thickness of the oxide layer between the lithium niobate and the electrode varies across the wafer due to the non-uniformity of the mechanical polishing rate. Gold electrodes were patterned with a direct laser writing tool to produce a nominal electrode gap of 3.5 µm. A double-layer metal lift-off was used to define the gold electrodes with a nominal thickness of 900 nm. Finally, the edges of the sample were mechanically polished for lensed fiber coupling to waveguides with polished end facets.

**Funding.** National Institute of Standards and Technology NIST-on-a-Chip program.

**Acknowledgements.** We thank J. R. Lawall for helpful discussions.

**Disclosures.** Related patent applications have been submitted.

**Data availability.** The data underlying this paper will be publicly available at a data.nist.gov.

**Supplemental Document.** See the Supplemental Materials for further discussion on the methods.